\documentclass[aps,pre,twocolumn,groupedaddress,showpacs]{revtex4}
\usepackage{graphicx}
\newcommand{\sep}{ \ \ \ , \ \ \ }

\newcommand{\beq}{\begin{equation}}
\newcommand{\eeq}{\end{equation}}
\newcommand{\beqn}{\begin{eqnarray}}
\newcommand{\eeqn}{\end{eqnarray}}

\newcommand{\dd}{{\rm d}}
\newcommand{\ee}{{\rm e}}
\newcommand{\eq}{Eq.\ }
\newcommand{\eqs}{Eqs }
\newcommand{\fig}{Fig.\ }

\newcommand{\cf}{cf.\ }
\newcommand{\ct}{c_{\rm tot}}

\newcommand{\Tab}{Table }
\newcommand{\app}{Appendix }
\newcommand{\De}{D_{\rm eff}}

\begin{document}
%\begin{CJK*}{CNS1}{}
%\preprint{}

\title{Isotropic-nematic phase transition in amyloid fibrilization
}
\author{Chiu Fan \surname{Lee}
% (\CJKchar[CNS1]{"4A}{"57} \CJKchar[CNS1]{"62}{"3E} \CJKchar[CNS1]{"47}{"7D})
%(\CJKchar[Bg5]{167}{245} \CJKchar[Bg5]{182}{087} \CJKchar[Bg5]{166}{124})
}
\email{C.Lee1@physics.ox.ac.uk}
\affiliation{Physics Department, Clarendon Laboratory,
Oxford University, Parks Road, Oxford OX1 3PU, UK}


\date{\today}

\begin{abstract}
We carry out a theoretical study on the isotropic-nematic phase transition and phase separation in %%@
amyloid fibril solutions. 
Borrowing the thermodynamic model employed in the study of cylindrical micelles, we investigate the %%@
variations in the fibril length distribution and phase behavior with respect to changes in the %%@
protein concentration, fibril's rigidity, and binding energy. We then relate our theoretical %%@
findings to the nematic ordering experimentally observed
in  Hen Lysozyme fibril solution.
\end{abstract}
%\keywords{}

\maketitle

%\end{CJK*}

\section{Introduction}

Amyloids are insoluble fibrous protein aggregations stabilized by a network of hydrogen bonds and %%@
hydrophobic interactions \cite{Sunde_JMB97, Dobson_Nature03, %%@
Radford_TrendsBiochemSci00,Sawaya_Nature07}.
They are intimately related to many neurodegenerative diseases such as the Alzheimer's Disease, the %%@
Parkinson Disease and other prion diseases \cite{Harper_AnnuRevBiochem97}.
Furthermore, it has recently emerged that 
 non-pathogenic amyloid fibrils possess great technological potential.  In particular, amyloid %%@
fibrils have been employed as nanowire templates \cite{Reches_Science03,Scheibel_PNAS03}, were %%@
shown to possess great tensile strength \cite{Knowles_Science07,Smith_PNAS06} and complex phase %%@
behavior similar to liquid crystals \cite{Aggeli_PNAS01,Sagis_Langmuir04,Corrigan_JACS06}.  Given %%@
these extraordinary properties, it is highly desirable to investigate how one may exploit amyloid %%@
fibrils as functional materials. Here, we study theoretically the isotropic-nematic phase %%@
transition in amyloid fibril solutions by combining the physics of self-assembled linear %%@
structures, as studied in cylindrical micelles (see \cite{Schoot_Langmuir94,Schoot_EPL94} and the %%@
references therein), and the physics of the nematic ordering in charged rods %%@
\cite{Stroobants_Macromol86,Vroege_RepProgPhys92}.  We then apply the formalism to a specific %%@
example -- Hen Lysozyme (HL) fibril solution, and discuss agreements between theory and the %%@
experimental results in \cite{Corrigan_JACS06}. 

In the next section, we introduce a toy model for amyloid fibrilization and review briefly the %%@
physics of nematic ordering in self-assembled rods. In Section III, we apply the theoretical %%@
formalism to Hen Lysozyme (HL) amyloid fibrils and estimate all of the model parameters from %%@
previous experimental studies. We then discuss the limitations of and predictions from the model in %%@
Section IV.

%%%%%%%%%%%%%%%%%%%%%%%%%%%%%%%%%%%%%%%%%%%%%%%%%%%%%%%%
\begin{figure}[b]
\caption{(Color online)  Schematic diagrams of a monomer and a fibril. (a) The monomers interact %%@
with each other via two types of directional interactions indicated by the blue (dark grey) arrows %%@
($A$-type interactions) and the red (shaded areas between the beads) patches ($B$-type %%@
interactions) (see text). 
(b) A fibril is formed by joining the blue arrows and red patches. The directionality of the %%@
$A$-type interactions renders the fibril rod-like, and the chiral nature of the beta strands %%@
restricts the number of filaments, $\gamma$, in a fibril \cite{Turner_PRL03}. In the fibril %%@
depicted below, 
$\gamma =2$.
}
\label{fibril}
\begin{center}
\includegraphics[scale=.4]{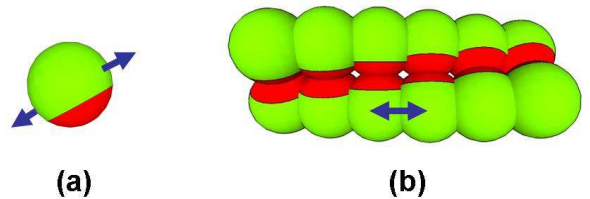}
\end{center}
\end{figure}
%%%%%%%%%%%%%%%%%%%%%%%%%%%%%%%%%%%%%%%%%%%%%%%%%%%%%%%

\section{The model}
%{\it The model.}
We assume that the monomers self-assemble into the fibrillar form through two different %%@
interactions: 
i) $A$-type interactions of strength $\alpha$ which are directed longitudinally along the fibrillar %%@
axis, and ii) $B$-type interactions of strength $\beta$ which are lateral to  the fibrillar axis.
For amyloid fibrils, the $A$-type interactions would correspond to the hydrogen bonds among the %%@
beta strands and hydrophobic interactions between the side chains packed between the beta sheets; %%@
and the $B$-type interactions would correspond to the  inter-cross-beta-sheet interactions %%@
\cite{Sunde_JMB97} (\cf \fig \ref{fibril}). We further assume that all fibrils are formed with the %%@
same number of filaments, $\gamma$, which is peptide specific and is constrained by the chiral %%@
nature of the cross-beta sheets \cite{Turner_PRL03,vanGestel_BiophysJ06}.

Given $N$ monomers in a volume $V$ of solution, we denote the fibrillar aggregate consisting of $s$ %%@
monomers by $N_s$, i.e., $\sum_s sN_s =N$. We consider only the fibrillar species and ignore the %%@
free energy contributions from monomers and oligomers not in the fibrillar form. This assumption is %%@
satisfied if the concentration is much higher than the Critical Fibrillar Concentration (CFC) (so %%@
that the monomer concentration is negligible), and if the CFC is lower than the critical %%@
concentrations of other oligomeric species \cite{Lee_a2008}. These conditions are met  if fibrils %%@
constitute the most dominant species in the solution.

As each monomer is a peptide, there are intrinsic internal degrees of freedom which 
contribute to the partition function.
This is equivalent to the contribution of configurational entropy for polymers. To 
simplify our theoretical treatment, we will absorb these degrees of freedom intrinsic to each %%@
monomer into $\beta$ 
and represent the monomer as a spherical 
particle.
 Since the free energy is defined up to addition of a constant, we will also set the %%@
monomer-solvent interaction energy to zero so that the free energy for free monomers becomes purely %%@
entropic. 

Without fibril-fibril interactions, the overall configurational partition function can be written %%@
as (\cf \app \ref{A_part}):
\beq
\label{partition}
Z_{\rm tot} = \prod_s' \frac{(Z_s)^{N_s}}{N_s!}
\eeq
where the prime in the product denotes the restriction that $\sum_s sN_s =N$, and
\beqn
\nonumber
Z_s &=&  \frac{\Psi^{s}(u \theta)^{s-1} }{\Lambda^{3s}}\exp[(s-\gamma)\alpha+s\beta]\ .
\eeqn
To ease notation, we define two new parameters:
\beqn
\label{chi}
\chi &\equiv &\alpha +\beta +\ln (u \theta) +\ln ( \Lambda^{-3} \Psi )
\\
\label{xi}
\xi &\equiv &\gamma \alpha +\ln (u \theta) \ .
\eeqn
Namely, $\chi$ amounts to the sum of the monomer binding energies through the $A$-type (first term) %%@
and $B$-type (second term) interactions, plus the entropic contribution (the third term) and the %%@
kinetic contribution (the fourth term); and $\xi$ amounts to the total longitudinal binding %%@
energies of the fibril, plus the entropic contribution.

According to \eq (\ref{partition}), the Free Energy Density (FED), in the absence of fibril-fibril %%@
interactions, 
is expressed as:
\beq
\label{f_0}
f_0 = \int \dd s n(s) \left[ \ln n(s)-\chi s+\xi -1  \right]
\eeq
where $k_BT$ is set to one and $n(s) \equiv N(s)/V$ with the unit volume set to be the volume of %%@
one monomer. Under this convention, $n(s)$ is dimensionless and corresponds to the volume fraction. 
Note that in \eq (\ref{f_0}), we have gone  from a discrete description of the aggregation number %%@
to a continuous one. This assumption is valid if the mean aggregation number is large.

To incorporate the steric interactions between fibrils, we employ the formalism developed in the %%@
study of cylindrical micelles \cite{Odijk_Macromol86,Schoot_EPL94}. Specifically, we model the free %%@
energy contribution of the fibril-fibril interactions as
\beq
 f_{\rm int} = \int \dd s \dd s' n(s) n(s') B(s,s')
 \eeq
 where $B(s,s')$ is the second virial coefficients of two rods of aggregation numbers $s$ and $s'$. %%@
Denoting the diameter of the fibril by $D$ and the length of a fibril with $s$ monomers by $L(s)$, %%@
we have \cite{Schoot_Langmuir94}:
 \beq
 B(s,s')=
\frac{2 \pi}{3} D^3 +\frac{\pi^2}{2} D^2 [L(s) +L(s')] +
DL( s)L(s') |\sin \phi | \ ,
\eeq
where $\phi$ is the angle between the two rods.
Since the mean fibrillar length is much greater than $D$ in our systems of interest, we will ignore %%@
the first two terms in the second virial coefficients.  Also, as a fibril is a linear structure, 
\beq
\label{zeta2}
L(s) = \zeta s \ ,
\eeq
for some constant $\zeta$. We shall from now on express the FED in terms of $\zeta$. 

In the isotropic phase, the different directions of the rods are averaged over and so $f_I$ is %%@
\cite{Schoot_Langmuir94}:
\beq
\label{fI}
f_I = f_0+\frac{\pi D \zeta^2}{4}\int \dd s \dd s' s s'n(s) n(s')  \  ,
\eeq
In the nematic phase, the flexibility of the fibrils has to be taken into account to avoid length %%@
explosion in the nematic phase \cite{Schoot_Langmuir94}. This can be done by incorporating the %%@
persistence length of the fibril, denoted by $P$, into the model. The resulting FED for the nematic %%@
phase is \cite{Schoot_EPL94}:
\beqn
\nonumber
f_N &=& f_0+\int \dd s\ n(s) \left[
\ln \frac{P}{4 \lambda} +\frac{\zeta s}{4  \lambda} \right]
\\
\label{fN}
&&+D\zeta^2 \sqrt{\frac{\lambda \pi}{P}} \int \dd s \dd s's s' n(s) n(s') \ ,
\eeqn
where $\lambda$ is the deflection length of the fibril \footnote{In brief, $\lambda$ corresponds to %%@
the length scale on which the fibril is deflected 
from its ideal path in order to conform to the nematic constraint (\cf \cite{Odijk_Macromol86}). }. %%@
In particular, $\lambda$ is related to the orientational order parameter, $\eta$, in the following %%@
manner \cite{Odijk_Macromol86}:
\beq
 \eta \simeq (1 - 3\lambda/P)\ .
 \eeq

The length distribution that minimizes the above FEDs can now be found by using the Lagrange %%@
multiplier method (e.g. see \cite{Schoot_Langmuir94,Schoot_EPL94}). For the isotropic phase, the %%@
distribution is :
\beqn
\label{nI}
n_I(s) &=&  \exp[-s/S_I -\xi]
\\
\label{SI}
 S_I &=& \sqrt{c_I\ee^\xi}  \ ,
\eeqn
where $c_I$ is the protein volume fraction and $S_I$ corresponds to the average aggregation number. %%@
Note that $S_I$ is above one only if $c_I > \ee^{-\xi}$, this therefore indicates that the CFC of %%@
the system is at $\ee^{-\xi}$ \cite{Lee_a2008}.

For the nematic phase, the distribution is
\beqn
\label{nN}
n_N(s) &=&  \frac{4\lambda^*}{P}\exp[-s/S_N -\xi]
\\
 S_N &=& \sqrt{\frac{Pc_N\ee^\xi}{4 \lambda^*}}
\eeqn
where, similar to the isotropic case, $c_N$ is the protein volume fraction and $S_N$ corresponds to %%@
the average aggregation number. In the above equations, $\lambda^*$ is determined by minimizing \eq %%@
(\ref{fN}) with respect to $\lambda$, and thus satisfies the following equation:
\beq
-\sqrt{\frac{c_N}{P\lambda^{*} \ee^\xi}}-\frac{\zeta c_N}{4  \lambda^{*2}}
+\frac{D\zeta^2}{2} \sqrt{\frac{\pi}{\lambda^* P}} c_N^2 =0\ .
\eeq
Due to the large magnitude of $\ee^\xi$ in the systems that we are interested in (\cf \Tab %%@
\ref{Ta1}), we will ignore the first term above and approximate $\lambda^*$ as:
\beq
\label{lambda}
\lambda^{*  3/2} = \frac{1}{2D \zeta c_N} \sqrt{\frac{P}{\pi}} \ .
\eeq
Substituting \eqs (\ref{nI}) and (\ref{nN}) into \eqs (\ref{fI}) and (\ref{fN}), the minimal FEDs %%@
for the two phases are:
\beqn
\label{fi}
f_I&=&-\left(\chi S_I^2 +2S_I \right)\ee^{-\xi} +\frac{\pi D \zeta^2c_I^2}{4}
\\
\nonumber
f_N &=&-\frac{4 \lambda^*}{P}\left[\left(\chi-\frac{\zeta}{4 \lambda^*}\right) S_N^2 %%@
+2S_N\right]\ee^{-\xi} 
\\
\label{fn}
&&+D\zeta^2 \sqrt{\frac{\lambda^* \pi}{P}}c_N^2 \ .
\eeqn
The FEDs above apply only to pure isotropic and nematic phases. To investigate the co-existence of %%@
the two phases, we denote by $v_I$ ($v_N$) the proportion of isotropic (nematic) component in the %%@
system. The total FED is therefore
\beq
\label{ftot}
f_{\rm tot}(c_I,c_N) = v_If_I(c_I)+ v_N f_N(c_N)\ ,
\eeq
with the following conditions:
\beqn
v_Ic_I + v_N c_N &=& c_{\rm tot}
\\
v_I +v_N &=&1 \ .
\eeqn
As $v_I$, $v_N$ can be expressed in terms of $c_I$, $c_N$:
\beq
v_I = \frac{c_N-\ct }{c_N -c_I} \sep v_N = \frac{\ct - c_I}{c_N -c_I}\ ,
\eeq
the total FED of the system is dependent only on $c_I$ and $c_N$. The proportion of the %%@
isotropic/nematic component can now be obtained by minimizing the total FED with respect to $c_I$ %%@
and $c_N$. This minimization problem does not admit an analytical expression and the graphical %%@
method described in  \cite{Doi_B86} is required (\cf \fig \ref{result1}). It is therefore %%@
worthwhile putting in experimentally relevant values into the model to reduce the number of  %%@
variables to be analyzed. This leads us to the next section where the HL fibril solution is %%@
discussed. But before we do so, we note that as far as the length distribution and phase behavior %%@
are concerned, the energy strength, $\chi$, is irrelevant. This is because the terms involving %%@
$\chi$ in the total FED, i.e., the first terms in \eqs (\ref{fi}) and (\ref{fn}), amounts to
\beq
-v_I \chi S_I^2 \ee^{-\xi} -v_N\frac{4 \lambda^*}{P}\chi S_N^2 \ee^{-\xi}
=- \chi \ee^{-\xi} \ ,
\eeq
which is independent of $c_I$ or $c_N$. One consequence of this realization is that one will not be %%@
able to obtain a full picture of the overall fibrilization energy, which includes the longitudinal %%@
binding term, $\alpha$, as well as the lateral binding term, $\beta$, by studying the length 
distribution and the phase behavior of the system alone.

%%%%%%%%%%%%%%%%%%%%%%%%%%%%%%%%%%%%%%%%%%%%%%%%%%%%%%%%
\begin{figure}
\caption{(Color online)
Four beta sheet strands are shown with the fibrillar axis along the vertical direction. The %%@
hydrogen bonds are schematically displayed by the rods connecting the two pairs of beta sheets. On %%@
average, the amino-acids within a beta strand are about 0.35 nm apart. According to this structural %%@
model, each amino acid occupies on average $(1 \times 0.35 \times 0.48)= 0.17$~nm$^3$ of space %%@
within the fibril.
}
\label{fibrils}
\begin{center}
\includegraphics[scale=.35]{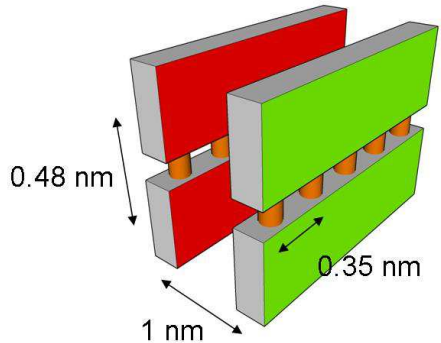}
\end{center}
\end{figure}
%%%%%%%%%%%%%%%%%%%%%%%%%%%%%%%%%%%%%%%%%%%%%%%%%%%%%%%

%%%%%%%%%%%%%%%%%%%%%%
\begin{table}
\caption{The various parameters involved in our model. 
The daggers indicate values predicted in this work.
}
\label{Ta1}
\begin{center}
\begin{tabular}{c|c|c} \hline 
Properties \ \ &\ \ Symbols\ \ &\ \ Hen Lysozyme\ \ 
\\ \hline 
No.\ of AA\ \ &  -- & 129 
\\
Persistence length & $P$ & 10 $\mu$m \cite{Knowles_Science07} 
\\
Diameter & $D$ & 7.4 nm \cite{Corrigan_JACS06}
\\
Effective diameter & $D_{\rm eff}$ & 13 nm
\\
Fibril length per monomer \ & $\zeta$ & 0.5 nm 
\\
Binding free energy  & $\xi$ & $20.6$$^\dagger$ 
\\
Lower conc.\ for phase sep.\ & $c_A$ & \ $\sim 0.65$ mM \cite{Corrigan_JACS06} 
\\
Upper conc.\ for phase sep.\ & $c_B$ & 1.05 mM$^\dagger$ 
\\
\hline 
\end{tabular}
\end{center}
\end{table}
%%%%%%%%%%%%%%%%%%%%%%%%%%%%%%

\section{Hen Lysozyme amyloid fibrils}
We will now apply our theoretical formalism to a specific system -- the HL fibril solution.
 HL is a 
protein consisting of 129 Amino Acids (AA) and amyloid fibrils are observed to form when incubated %%@
at low pH and elevated temperatures \cite{Krebs_JMB00,Corrigan_JACS06}. The width and persistence %%@
length of a HL fibril is found to be 7.4 nm \cite{Krebs_JMB00,Corrigan_JACS06} and 10 $\mu$m %%@
\cite{Knowles_Science07}, respectively. To complete the list of parameters involved in the model, %%@
we need to estimate $\zeta$ (\cf \eq (\ref{zeta2})) and $\xi$ (\cf \eq (\ref{xi})). Since the %%@
structural details of the HL fibrils are still lacking, we will estimate $\zeta$ by employing the %%@
approximation adopted in \cite{Knowles_Science07} in the study of elastic properties of amyloid %%@
fibrils.
Specifically, each amino acid is assumed to occupy a volume of $(1 \times 0.48  \times 0.35) = %%@
0.17$ nm$^3$ within the fibril. This assumption is motivated by the fact that a fibril constitutes %%@
mainly of cross-beta-sheet structure (\cf \fig \ref{fibrils}).
Given that $D = 7.4$ nm \cite{Corrigan_JACS06}, the average length contribution to the fibril per %%@
monomer can be estimated as follows:
\beq
\label{zeta}
\zeta = \frac{129 \times 0.17 \ {\rm nm}^3}{( D/2)^2 \pi } = 0.5\ {\rm nm} .
\eeq
In other words, each HL protein in the fibrils contributes on average 0.5 nm to the fibril's %%@
length. 

When the solution is at pH 2 in the presence of 100 mM NaCl, a HL protein carries a net positive %%@
charge of 19 \cite{Corrigan_JACS06}. Hence, a HL fibril has an average line charge density, $\nu$, %%@
of $(19 / \zeta) = 38$/nm. The electrostatic repulsion due to the fibrils' charge density can be %%@
accounted for by defining an effective diameter for the fibril, which is of the form \footnote{Note %%@
that besides the effective change in the fibrils' diameter due to the charges, there should also be %%@
a twist term added to the FED due to the preferred perpendicular configuration among fibrils %%@
\cite{Stroobants_Macromol86, Vroege_RepProgPhys92}. This term is ignored here due to its small %%@
magnitude here.}:
\beq
\label{De}
\De = D\left[1+\frac{\ln A +0.577 +\ln 2 -1/2}{\kappa D} \right]
\eeq
where 
\beqn
\label{Debye}
\kappa^{-1} &=&  \sqrt{\frac{\epsilon_0 \epsilon k_BT}{2 N_A e^2 I}} \simeq 1~{\rm nm}
\\
Q &=& \frac{e^2}{4 \pi \epsilon_0 \epsilon k_BT} \simeq 0.68~{\rm nm}
\\
\label{AA}
A &= & \frac{\pi \Gamma^2 }{2 Q \kappa\ee^{\kappa D}} \simeq 114 
 \ .
\eeqn
In \eq (\ref{AA}), $\Gamma \simeq 300$, and is estimated by solving the Poisson-Boltzmann equation %%@
for a charged cylinder in an ionic solvent (\cf \app \ref{Gamma}). Note that in the above %%@
equations, $\epsilon_0$, $\epsilon$, $N_A$, $e$, $I$, $\kappa^{-1}$, and $Q$ are the vacuum %%@
permittivity, the dielectric permittivity of the solvent (taken to be 82 here), the Avogadro %%@
number, the elementary charge, the ionic strength of the solvent in unit of mole/m$^3$, the Debye %%@
screening length, and the Bjerrum length, respectively \cite{Stroobants_Macromol86, %%@
Vroege_RepProgPhys92}. Substituting the values in 
 \eqs (\ref{Debye}) and (\ref{AA}) to \eq (\ref{De}),  $\De$ can be calculated to be about $1.7 %%@
\times D \simeq 13$ nm. We will employ this effective diameter in our FEDs shown in 
\eqs (\ref{fi}) and (\ref{fn}).

\begin{figure}
\caption{(Color online) 
The plot of $f(c) +\chi c$ vs.\ $c$ for the isotropic and nematic phases. The dotted line is %%@
tangent to both curves and the contact points (indicated by the arrows) correspond to the lower %%@
concentration, $c_A = 0.65$ mM, and upper concentration, $c_B = 1.05$ mM, for phase separation %%@
\cite{Doi_B86}. 
}
\label{result1}
\begin{center}
\includegraphics[scale=.45]{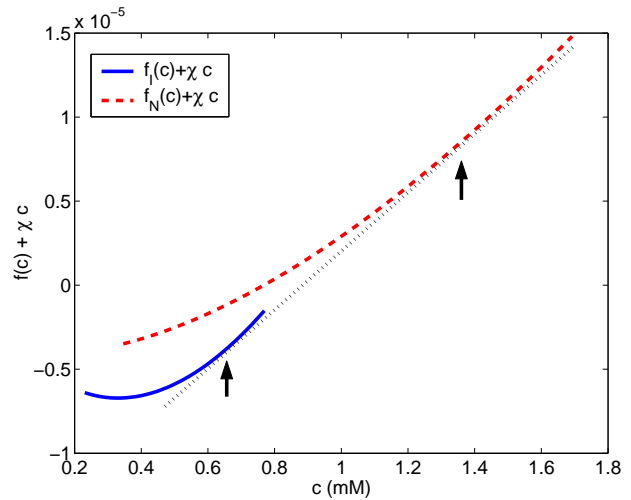}
\end{center}
\end{figure}

\begin{figure}
\caption{The plots of a) nematic volume fraction, b) orientational order parameter, and c) average %%@
fibril length vs.\ HL concentration. 
}
\label{result2}
\begin{center}
\includegraphics[scale=.45]{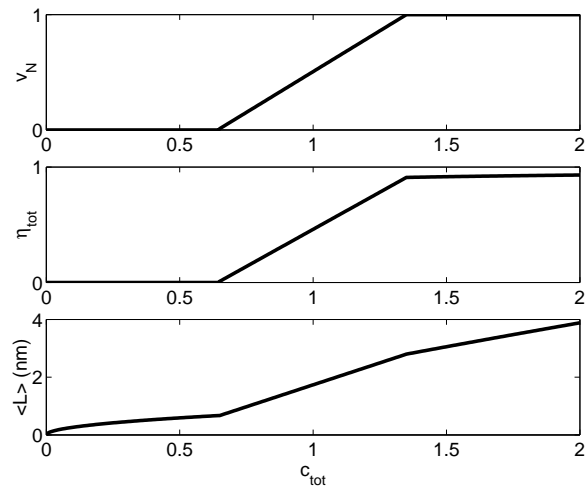}
\end{center}
\end{figure}

To determine the only remaining parameter $\xi$, we will make use of the knowledge that the lower %%@
concentration for phase separation, $c_A$, is measured to be around 0.6--0.7 mM %%@
\cite{Corrigan_JACS06}. Here, we set $c_A$ to be 0.65 mM for definiteness. We then vary $\xi$ until %%@
$c_A$ obtained from the tangent method (illustrated in \fig \ref{result1})  matches the assigned %%@
value of 0.65 mM.  In doing so, we find that $\xi \simeq 20.6$ and the upper concentration for %%@
phase separation, $c_B$, is about 1.05 mM. 
We have now completely specified all of the model parameters, and based on these parameters, the %%@
variations of the various properties of the HL fibril solution with respect to protein %%@
concentration is shown in \fig \ref{result2}. 

%%%%%%%%%%%%%%%%%%%%%%%%%%%%%%%%%%%%%%%%%%%%%%%%%%%%%%%%
%%%%%%%%%%%%%%%%%%%%%%%%%%%%%%%%%%%%%%%%%%%%%%%%%%%%%%%

%%%%%%%%%%%%%%%%%%%%%%%%%%%%%%%%%%%%%%%%%%%%%%%%%%%%%%%%
%%%%%%%%%%%%%%%%%%%%%%%%%%%%%%%%%%%%%%%%%%%%%%%%%%%%%%%

\section{Discussion}
Starting from a toy thermodynamic model , we have studied the isotropic-nematic phase transition in %%@
amyloid fibril solution by combining previous models for  i) the nematic ordering in self-assembled %%@
linear structures, ii) the nematic ordering in charged rods, and iii) the elastic properties of %%@
amyloid fibrils. We then focus on HL fibril solution and estimated  
all of the parameters involved from experimental values. 
From the resulting parameter-free model, we deduced two main predictions: i) the upper %%@
concentration for phase separation, $c_B$, is 1.05 mM; and ii) the average fibril length varies %%@
with protein concentration in the way depicted in \fig \ref{result2}(c). In particular, the average %%@
fibril length is predicted to be 1.21 $\mu$m at the protein concentration of 0.5 mM.  The first %%@
prediction on the upper concentration seems to be an underestimate of the upper concentration %%@
observed experimentally in \cite{Corrigan_JACS06}. This may be an outcome of further aggregation of %%@
fibrils. Indeed, it has been shown that amyloid fibril solutions tend to form gel at high %%@
concentration \cite{Sagis_Langmuir04, Corrigan_EPJE09}. Gelation in fibril solutions points to %%@
possible attractive interactions between fibrils \cite{Philipse_Langmuir98}, or contacts induced %%@
viscoelasticity \cite{Philipse_Langmuir96}. 
These effects are not captured in our FEDs, and may be the source of the discrepancy. 
On the prediction concerning the average fibril length, to the best of our knowledge, the average %%@
fibril length for HL fibrils has not been determined accurately and so this prediction remained to %%@
be verified. 

Besides protein concentration, salt concentration is also demonstrated to have a major effect on %%@
the onset of nematic ordering in \cite{Corrigan_JACS06}. Within our model, a decrease in ionic %%@
strength would increase the Debye screening length (\cf \eq (\ref{Debye})), and hence also the %%@
effective diameter. This implies that a decrease in ionic strength would decrease the onset %%@
concentration for nematic ordering, which is indeed observed experimentally in %%@
\cite{Corrigan_JACS06}. 

\begin{acknowledgements}
The author thanks the Glasstone Trust (Oxford) and Jesus College (Oxford) for support.
\end{acknowledgements}

\appendix
\section{The Partition function}
\label{A_part}
The partition function for a fibril  consisting of $s$ monomers is
\beq
\label{Qs2}
Z_s = \frac{\Psi^s}{\Lambda^{3s} s!}\int_{\Gamma_s} \dd x_1 \cdots \dd x_s \dd w_1 \cdots \dd w_s %%@
\ee^{-U(\{ x,w\}) / k_BT}
\eeq
where $x$ ($w$) are the coordinates (directors) for the monomers, 
$U(\{ x,w\})$ is the potential function, and $\Gamma_s$ constrains the positions and directions of %%@
the monomers so that the aggregation is in the fibrillar form. Note that the prefactor corresponds %%@
to the kinetic part of the partition function such that 
$\Lambda =h/\sqrt{2\pi m k_BT}$ is the de Broglie thermal wavelength, and $\Psi = \sqrt{ (2\pi)^5 %%@
(k_BT)^3 I_1I_2I_3}/h^3$ with $I_i$ being the three principle moments of inertia %%@
\cite{Mutaftschiev_B01}. 

In the mean-field limit where all monomer contributions to the partition function are assumed to be  %%@
identical, the above integral can be partitioned into four terms \cite{Mutaftschiev_B01}:
\beq
\label{four}
 \underbrace{\frac{\Psi^s}{\Lambda^{3s} }}_{\rm kinetics} \times \underbrace{V u^{s-1}}_{\rm %%@
translation}\times \underbrace{4\pi\theta^{s-1}}_{\rm rotation} \times \underbrace{\ee^{E_s}}_{\rm %%@
binding}
\ .
\eeq
The first term corresponds to the kinetic contribution, the second term to the translational %%@
entropic contribution with $u$ being the roaming volume of each monomer within the fibril, the %%@
third term to the rotational entropic contribution with $\theta$ being the roaming area on a unit %%@
sphere for the director of each monomer, and the fourth term to the binding energy, which is of the %%@
form:
\beq
E_s = (s-\gamma) \alpha+s \beta \ .
\eeq
In the above equation, $\alpha$ ($\beta$) is the binding energy corresponding to the $A$-type %%@
($B$-type) interactions (\cf \fig \ref{fibril}). Note also that the term $s!$ in the denominator in %%@
\eq (\ref{Qs2}) disappeared in \eq (\ref{four}) due to the fact that there are $s!$ different ways %%@
of shuffling the monomers within the fibril.

For a system with fibrils of variable lengths, we need to sum over all of the partition functions %%@
for the $s$-fibril, the total partition function is therefore:
\beq
Z_{\rm tot} = \prod_s' \frac{(Z_s)^{N_s}}{N_s!}
\eeq
where $N_s$ is the number of $s$-fibrils and the prime in the product denotes the restriction that %%@
$\sum_s sN_s =N$.

\section{Estimation for $\Gamma$}
\label{Gamma}
The electrostatic potential $\psi$ for a long, cylindrical and charged rod in a solution with %%@
excess salt of ionic strength $I$ is described by \cite{Philip_JChemPhys70}:
\beq
\label{psi1}
\frac{1}{r} \frac{\dd}{\dd r} \left( r \frac{ \dd \psi}{\dd r} \right) = \frac{8 \pi N_A I %%@
e^2}{\epsilon_0 \epsilon} \sinh \frac{e \psi}{k_BT}
\eeq
with the boundary conditions:
\beqn
\label{B1}
\left. \frac{\dd \psi}{\dd r} \right|_{r = D} &=& -\frac{4 \nu}{\epsilon_0 \epsilon D}
\\
\lim_{r \rightarrow \infty} \psi (r) &=& 0 \ ,
\eeqn
where $r$ denotes the distance away from the center line of the fibril. Note that we have assumed %%@
in \eq (\ref{B1}) that all of the charges are spread at the outer boundary of the cylinder.

At a distance $r$ far away from the rod $(\kappa r > \kappa D /2 +1)$,
\beq
\frac{e \psi(r)}{k_BT} \simeq \Gamma K_0(\kappa r) 
\eeq
where $K_0(.)$ is the modified Bessel function of the second kind \cite{Philip_JChemPhys70}. The %%@
prefactor $\Gamma$ can therefore be obtained by solving the differential equation \eq (\ref{psi1}) %%@
\cite{Philip_JChemPhys70}.

%\bibliography{chiufanlee}

\end{document}